\documentstyle[epsf,psfig]{article}
\newcommand{\nIs}{n_{1\s}}
\newcommand{\nJs}{n_{2\s}}

\newcommand{\R}{\Rv}
\newcommand{\Rv}{\bf{R}}
\newcommand{\rv}{\bf{r}}

\newcommand{\beq}{\begin{equation}}
\newcommand{\eeq}{\end{equation}}
\newcommand{\s}{\sigma}
\newcommand{\ra}{\rangle}
\newcommand{\la}{\langle}
\newcommand{\rta}{\rightarrow}

\font\nineit=cmti9
\font\ninebf=cmbx9

\begin{document}

\centerline{\bf DISTANCE-DEPENDING ELECTRON-PHONON INTERACTIONS}
\vspace*{0.035truein}
\centerline{\bf  FROM ONE- AND TWO-BODY ELECTRONIC TERMS IN A DIMER.}

\vspace*{0.37truein}
\centerline{M. ACQUARONE}
\vspace*{0.015truein}
\centerline{\it C.N.R.-G.N.S.M., Unita' I.N.F.M., Dipartimento di Fisica,Universita' di Parma}
\baselineskip=10pt
\centerline{\it Parma, I-43100, Italy}
\vspace*{10pt}
\centerline{ and}
\vspace*{10pt}
\centerline{C. NOCE}
\vspace*{0.015truein}
\centerline{\it Unit\`a I.N.F.M.,
Dipartimento di Scienze Fisiche ''E.R. Caianiello'', Universit\`a di Salerno,}
\baselineskip=10pt
\centerline{\it Baronissi, I-84081, Italy}
\vspace*{0.225truein}

\vspace*{0.21truein}
\centerline{Abstract.}
For a dimer with a non-degenerate orbital built
from atomic wave functions of Gaussian shape we
evaluate all the  electron-phonon couplings derived from the one-body
and two-body electronic interactions, considering
 both the adiabatic and extreme non-adiabatic
limit. Not only the  values
 of the coupling parameters in the two limits, but also the 
 expressions of the corresponding terms in the
Hamiltonian differ.
 Depending on the distance between the dimer ions,
some of the two-body couplings are comparable, or even larger than
the one-body ones

\vspace*{12pt}			

\vspace*{1pt}	
\section{The Model.}	
\vspace*{-0.5pt}
\noindent
 In a general two-site electron-phonon Hamiltonian
$H=H_{el}+H_{ph}+H_{el-ph}  \label{ham1}$ the interacting term
$H_{el-ph}$ originates from developing $H_{el}$ to first order in the ion displacements 
$u_i\quad (i=1,2)$, where, in standard notation for a non-degenerate orbital:
\[
H_{el}=\epsilon \sum_{\sigma }(n_{1\sigma }+n_{2\sigma})+\sum_{\sigma}
[t+X(n_{1-\sigma }+n_{2-\sigma})](c_{1\sigma }^{\dagger }c_{2\sigma }+H.c.)
+U(n_{1\uparrow}n_{1\downarrow}+n_{2\uparrow}n_{2\downarrow})
\]
\begin{equation}
+(V-J/2)n_1n_2-2J\left[S^z_1S^z_2+{{1}\over{2}}(S^+_1S^-_2+H.c.)\right]
+P(c^{\dagger}_{1\uparrow}c^{\dagger}_{1\downarrow}
c^{}_{2\downarrow}c^{}_{2\uparrow}+H.c.).
 \label{helbare}
\end{equation}
 We shall develop both the one-($\epsilon,t$) and two-body($X,U,V,J_z=J_{xy}=P$)
 electron interaction parameters, evaluated as in Ref.1, by assuming
 a non-degenerate orbital described
by Wannier functions built from atomic orbitals of Gaussian 
shape. We associate to each site a Gaussian atomic-like orbital
$\phi_i(\rv-\R_i)$, with the ions centered at  the positions
$\R_i\equiv(\pm a/2+u_i,0,0)\quad (i=1,2)$ . By defining
$N\equiv\left (2/ \pi\right)^{3/4}\Gamma^{3/2}$, they read:
$\phi_i(\rv)= N\exp\left\{-\Gamma^2\left[\left(x\pm
a/2-u_i\right)^2+y^2+z^2\right]\right\}$.
 Then the Wannier functions
$\Psi_1, \Psi_2$ can be written as:
\[
\Psi_1(\rv)=A(S) \phi_1(\rv)+ B(S) \phi_2(\rv) \qquad
\Psi_2(\rv)=B(S) \phi_1(\rv) +A(S)\phi_2(\rv)
\]
\beq 
A(S)=[(1+S)^{-1/2}+(1-S)^{-1/2}]/2 \qquad B(S)=[(1+S)^{-1/2}-(1-S)^{-1/2}]/2
\eeq
 with $S(u) \equiv \la \phi_1|\phi_2\ra =\exp[-\Gamma^2(a+u)^2/2]$, and $u=u_2-u_1$.
 To make clear our method of calculation, it is convenient to explicitate, as an example,
 the one-body local electronic energy:
\beq
\epsilon^{(i)}=\int\Psi_i^{\ast}(\rv,\R_1,\R_2)
\left[-{{\hbar^2}\over{2m}}\nabla^2+V_1(\rv-\R_1)+V_2(\rv-\R_2)\right]
\Psi_i^{~}(\rv,\R_1,\R_2) d^3\rv
\eeq
 where the potentials originating from the ion cores at
the displaced positions $\R_1$ and $\R_2$ are:
\beq
V_i\equiv V(\rv-\R_i)=-e^2Z\left[\left(x\pm a/2- u_i
\right)^2+y^2+z^2\right]^{-1/2}\qquad (i=1,2)
\eeq
with $-e$ the electron charge, and $+Ze$ the charge of the ion core. 
The local energy  $\epsilon$ (actually site-independent) can be
decomposed into three terms, corresponding to the contributions
from the kinetic operator ($\epsilon_{\nabla}^{(i)}$) and from
each one of the  ionic potentials
($\epsilon_{V_1}^{(i)},\epsilon_{V_2}^{(i)}$, respectively).
 A similar decomposition holds for the hopping amplitude $t$. \par
We need to  distinguish between the adiabatic 
 and the non-adiabatic 
limit in evaluating the electron-phonon interactions, because the
integrals have different kernels in the two cases. Indeed, in the
adiabatic limit the  displacements affect both the potentials $ V(\rv-\R_i)$ and
the  Wannier functions $\Psi_i^{~}(\rv,\R_1,\R_2)$ , expressing the requisite that the
electronic charge distribution adjusts itself instantaneously at the
position of the ions.
We shall schematize the opposite situation, where the electrons are
slower than the ions, as realized by  the
electronic charge distribution staying centred around the undisplaced
ion position, while the potentials are centred on the displaced ions.
 We shall call this the extreme anti-adiabatic limit. In the literature\cite{ashkenazi}
 the two limits are also named from, respectively, Fr\"{o}lich and Bloch.\par

\textheight=7.8truein

\section{Coupling terms in the anti-adiabatic limit.}
\noindent
In the anti-adiabatic limit  $\epsilon^{~}_{\nabla}$ does not
  change, therefore no electron-phonon coupling originates from it.
The coupling terms derived from the two-body interactions are also identically
vanishing in this limit, because they involve the Wannier
functions and the inter-electronic Coulomb potential
 which are both insensitive to the displacements of the ions.
 The only non-vanishing electron-phonon non-adiabatic interactions  arise
from the variation of  the potential contributions to $\epsilon$ and
$t$. Let us now succinctly describe their evaluation. Full details are given in Ref.3.\par
 The perturbation of $\epsilon^{(i)}_{V_i}$ originates a term in the Hamiltonian
connecting the local charge with the local deformation:
$\sum_\s (g_0^{(1)}\nIs u_1+g_0^{(2)}\nJs u_2)$ . It has  a formal similarity with the Holstein 
coupling term\cite{holstein}, but it has a physically different origin, as Holstein\cite{holstein}
 considered electrons moving along  a chain of fixed spacing with vibrating diatomic molecules at its nodes. 
On site 1, its explicit evaluation\cite{macanio} yields:
\beq
g_0^{(1)}={{2\Gamma\sqrt{2/\pi} } \over {a}} \left\{B^2
\left[F_0(2a^2\Gamma^2)  -S^4\right] +4ABS \left[
F_0\left({{a^2\Gamma^2}\over{2}}\right)-S \right] \right\}.
\label{hol27c} \eeq

\noindent where $F_0(x)=x^{-1/2}Erf(x^{1/2})$. As, under site permutation,
$a\rightarrow -a$,  $g_0^{(1)}=-g_0^{(2)}$, as might have been anticipated  by considering that, for equal
charges $n_1=n_2$ and displacement amplitudes, with ${\bf e}_{12}\equiv (\Rv_2-\Rv_1)/a$, the energies
 $g_0^{(1)}n_1{\bf u}_1\bullet{\bf e}_{12}$
 and $g_0^{(2)}n_2{\bf u}_2\bullet{\bf e}_{12} $
on both sites  coincide. Now symmetry requires $u_1=-u_2$ (a constraint which does not hold for the 
original Holstein model\cite{holstein}) from which $g_0^{(2)}=-g_0^{(1)}$
follows. The contribution to the  Hamiltonian then reads:
$g_0^{(1)}\sum_{\sigma}(n_{1\sigma}^{~}u_1^{~}-n_{2\sigma}^{~}u_2^{~})$

The ``crystal-field'' coupling term $g_{cf}^{(ij)}$ expresses the change in the energy
$\epsilon_{V_j}^{(i)}$ .
To establish the form of this term, let us consider site $1$,
 with charge $n_{1}$.
Its energy, after a displacement ${\bf u}_{2}$ of the ion on site $2$,
changes by an amount
\nobreak{$E^{(1)}=g_{cf}^{(12)}n_{1}$  ${\bf u}_{2}\bullet {\bf e}_{12}$}. \ This can
be considered as the quantity measured by an observer sitting on ion $1$ and
watching the ion $2$ \ moved by ${\bf u}_{2}$ .\ The equivalent measurement
done by  an observer \ on ion $2$ watching the ion $1$ displaced by ${\bf u}%
_{1}$,\ yields $E^{(2)}=g_{cf}^{(21)}n_{2}{\bf u}_{1}\bullet {\bf
e}_{12}$.  Assuming $n_1=n_2,and u_1=-u_2$
one must have  $E^{(1)}=E^{(2)}$, implying $g_{cf}^{(12)}=-g_{cf}^{(21)}$.
 Therefore for the dimer as a whole one writes this term as
$g_{cf}^{{12}}\sum_{\sigma}(n_{1\sigma}u_{2}-n_{2\sigma}u_{1})$.
The explicit evaluation\cite{macanio} for site $1$ yields:

\beq
g^{(12)}_{cf}=-2A\left({{\Gamma}\over{a}}\right)\sqrt{{{2}\over{\pi}}}
\left[AF_0(2a^2\Gamma^2)+4BSF_0(a^2\Gamma^2/2)-4BS^2-AS^4\right].
\label{g1cfnadexpl} \eeq
  From Eq.\ref{g1cfnadexpl} and the change of sign of $a$ under site permutation, 
 $g^{(12)}=-g^{(21)}$ follows.\par
The Su-Schrieffer-Heeger\cite{SSH} ($SSH$) interaction $\gamma^{(12)}$, characterizing
 the $SSH$ Hamiltonian $ H_{SSH}\equiv \gamma^{(12)}\sum_\sigma (c_{1\sigma}^\dagger
c^{~}_{2\sigma}+c_{2\sigma}^\dagger
c^{~}_{1\sigma})(u_2-u_1)$, 
 is due to the modulation of the hopping amplitude $t$.
To preserve the invariance of $H_{SSH}$
under site permutation, $\gamma^{(12)}=-\gamma^{(21)}$ (see e.g refs.6 and 7).
Indeed, its explicit expression\cite{macanio} is:
\[
\gamma^{(12)}=4\sqrt{{{2}\over{\pi}}}\left({{\Gamma}\over{a}}\right)
\left\{{{AB}\over{2}}\left[S_0^4-(1-4a^2\Gamma^2)F_0(2a^2\Gamma^2)\right]
\right\}
\]
\beq
+4\sqrt{{{2}\over{\pi}}}\left({{\Gamma}\over{a}}\right)
\left\{(A^2+B^2)S_0\left[S_0-F_0(a^2\Gamma^2/2)\right]\right\}.
\label{ga12nadexpl}
\eeq
Under site permutation $a\rta -a$ and
$A\rta B$ so that $\gamma^{(12)}=-\gamma^{(21)}$ as expected.

In conclusion, in the non-adiabatic limit the complete electron-phonon
Hamiltonian is given by:

\[
H^{na}_{ep}= g_0^{(1)}\sum_\s (\nIs u_1-\nJs u_2)+
g_{cf}^{(12)}\sum_{\sigma}(n_{1\sigma}u_{2}-n_{2\sigma}u_{1})
\]
\beq +\gamma^{(12)}\sum_\sigma (c_{1\sigma}^\dagger
c^{~}_{2\sigma}+c_{2\sigma}^\dagger c^{~}_{1\sigma})(u_2-u_1)
\label{Hnonad}
\eeq

\section{Coupling terms in the adiabatic limit.}
\noindent
 In the explicit expression of the different electronic
 interactions in the adiabatically displaced state, $u$
invariably enters in the combination $a+u$. So, to obtain the
corresponding couplings,
 one can simply take the derivative with respect to $a$ of the
parameters in Eq.\ref{helbare} as evaluated in Ref.1. \par

There is some confusion in the literature about the correct form
of the electron-phonon Hamiltonian obtained, in the adiabatic limit, from
the variation of the local energy $\epsilon$, therefore we shall
devote some space to clarify this point. In this 
limit, $u_i\ne 0$ enters  both the charge distributions and  the potentials.
 The `` crystal field'' interaction couples the
charge on site $i$ to the {\it relative} position of site $j$ through the
modification of both the kinetic and the potential contributions to $\epsilon^{(i)}$.
 Let's place the origin of the $x$-coordinate 
onto one of the displaced ions, at $\Rv_i$, say. One has then to take
 into account the {\it relative} displacement of the ions.
 In the adiabatic limit, therefore, the
overall $\epsilon$-derived electron-phonon coupling term in the
Hamiltonian is: $g_{\epsilon}^{(12)}\sum_\s n_{1\s}(u_2-u_1)
+g_{\epsilon}^{(21)}\sum_\s n_{2\s}(u_1-u_2)$.
 As $g_{\epsilon}^{(12)}=-g_{\epsilon}^{(21)}$
 we can  write the total adiabatic
contribution from local energy terms to the electron-phonon
Hamiltonian as $H^{\epsilon}_{ep}=
g_{\epsilon}^{(12)}\sum_\s(\nJs+\nIs)(u_2-u_1)$. A Hamiltonian of this form was used in the papers of
  Ref.8, while those of Ref.9 proposed Hamiltonians incompatible with our results. 
After including the $SSH$ term, 
the complete one-body electron-phonon Hamiltonian in the adiabatic limit has therefore
the form\cite{noi2}:\par

\beq
H^{ad}_{ep}=
g_{\epsilon}^{(12)}\sum_{\sigma}(n_{1\sigma}+n_{2\sigma})(u_{2}-u_{1})
 +\gamma^{(12)}\sum_\sigma (c_{1\sigma}^\dagger
c^{~}_{2\sigma}+c_{2\sigma}^\dagger c^{~}_{1\sigma})(u_2-u_1).
\label{Hadiab}
\eeq

Coming now to the explicit expressions of the one-body coupling parameters,
let us write for conveniency
 $g^{(12)}_{\epsilon}\equiv g_{\nabla}^{(12)}+ g_{V}^{(12)}$
and $\gamma^{(12)}\equiv
\gamma^{(12)}_{\nabla}+\gamma^{(12)}_{V}$. We obtain:

\[
 g^{(12)}_{\nabla}=
-{{\hbar^2}\over{2m}}\left[{{a\Gamma^4S^2}\over{(1-S^2)^2}}\right]
\left[2\left(1-a^2\Gamma^2-S^2\right)\right]
\]
\[
 g^{(12)}_{V}=-Ze^2\left(2\Gamma\sqrt{{{2}\over{\pi}}}\right)
\left[{{\partial (A^2+B^2)}\over{\partial u}}+{{4ABS^2+(A^2+B^2)S^4}\over{a}}\right]
\]
\beq
-Ze^2\left(2\Gamma\sqrt{{{2}\over{\pi}}}\right)
\left\{F_0(2a^2\Gamma^2)\left[{{\partial (A^2+B^2)}\over{\partial u}}-
{{(A^2+B^2)}\over{a}}\right]+
(4S)\left[{{\partial AB}\over{\partial u}}
-{{AB}\over{a}}\left(1+a^2\Gamma^2)\right)\right]\right\}.
 \label{geps}
\eeq

\[
\gamma^{(12)}_{\nabla}=
{{\hbar^2}\over{2m}}\left[{{aS\Gamma^4}\over{(1-S^2)^2}}\right]
\left[2(1-S^2)-a^2\Gamma^2(1+S^2)\right]
\]
\[
\gamma^{(12)}_{V}=
-Ze^2\left(4\Gamma\sqrt{{{2}\over{\pi}}}\right)\Bigg\{
\left[{{\partial AB}\over{\partial u}}
+{{A^2+B^2}\over{a}}S^2+{{AB}\over{a}}S^4\right] +
\left[{{\partial AB}\over{\partial u}}
-{{AB}\over{a}}\right]F_0(2a^2\Gamma^2)
\]
\begin{equation}
+S\left[{{\partial (A^2+B^2)}\over{\partial u}}
-{{A^2+B^2}\over{a}}(1+a^2\Gamma^2)\right]F_0(a^2\Gamma^2/2)\Bigg\}.
\label{SSHV}
\end{equation}
Notice that, as the partial derivatives are linear in $a$, they
 change sign under site permutation.
\par
Also the two-body electronic interactions $U,V,J(=P),X$ of Eq. (1)
 give rise to electron-phonon couplings, all of the form:
\beq 
H_{Y}^{ep}= \left({{dY}\over{da}}\right) F(c^{\dagger}_{i\sigma},c^{~}_{j\sigma})
(u_j-u_i)\quad (i,j=1,2)
\eeq
 where $Y=U,V,X,J$ and
$F(c^{\dagger}_{i\sigma},c^{~}_{j\sigma})$ is the
function of Fermi operators representing the two-body interaction
whose amplitude is $Y$. From the results of Ref.1, their
evaluation is trivial. Notice that, as
$U(a)=J(a)+e^2\Gamma/\sqrt{\pi}$, then $dU/da=dJ/da=dP/da$. We list below their explicit expressions:
\[
{{dX}\over{da}}=-e^2{{\Gamma}\over{\sqrt{\pi}}}
\left[\left(-a\Gamma S\right)
 \frac{\left( 1+3S^{2}\right) }{(1-S^{2})^{3}}\right]
\left[1+2S^{2}+F_{0}\left( a^{2}\Gamma ^{2}\right)
 -2(1+S^{2})F_{0}\left( \frac{a^{2}\Gamma ^{2}}{4}\right) \right]
\]
\[
-e^{2}\frac{\Gamma }{\sqrt{\pi }}\left[ \frac{S/a}{(1-S^{2})^{2}}\right]
\left\{ 4a^{2}\Gamma ^{2}S^{2}\left[ F_{0}\left( \frac{a^{2}\Gamma ^{2}}{4}%
\right) -1\right] +S^{2}-F_{0}\left( a^{2}\Gamma ^{2}\right) \right\}
\]
\begin{equation}
-e^{2}\frac{\Gamma }{\sqrt{\pi }}\left[
\frac{S/a}{(1-S^{2})^{2}}\right] \left\{ 2(1+S^{2}) \left[
F_{0}\left( \frac{a^{2}\Gamma ^{2}}{4}\right) -\sqrt{S}\right]
\right\},   \label{dXda}
\end{equation}
\[
\frac{dU}{da}=
e^{2}\frac{\Gamma }{\sqrt{\pi }}\left[ \frac{-4a\Gamma
^{2}S^{2}}{(1-S^{2})^{3}}\right] \left[ 2-S^{2}+2S^{4}+S^{2}F_{0}\left(
a^{2}\Gamma ^{2}\right) -4S^{2}F_{0}\left( \frac{a^{2}\Gamma ^{2}}{4}\right)
\right]
\]
\[
+e^{2}\frac{\Gamma }{\sqrt{\pi }}\left[ \frac{S^{2}/a}{(1-S^{2})^{2}}\right] %
\Bigg\{2a^{2}\Gamma ^{2}\left[ 1-4S^{2}-F_{0}\left( a^{2}\Gamma ^{2}\right)
+4F_{0}\left( \frac{a^{2}\Gamma ^{2}}{4}\right) \right]
\]
\begin{equation}
+S^{2}-F_{0}\left( a^{2}\Gamma ^{2}\right) +4\left[ F_{0}\left( \frac{%
a^{2}\Gamma ^{2}}{4}\right) -\sqrt{S}\right] \Bigg\},
\label{dUda}
\end{equation}

\[
\frac{d(V-J_z/2)}{da}=
\]
\[
e^{2}\frac{\Gamma }{\sqrt{\pi }}\left[- \frac{4a\Gamma
^{2}S^{2}}{(1-S^{2})^{3}}\right] \left[ 3-S^2-8S^4-(7-5S^2)F_0(a^2\Gamma^2)
-4(1-3S^2)F_0\left({{a^2\Gamma^2}\over{4}}\right)\right]
\]
\[
+e^{2}\frac{\Gamma }{\sqrt{\pi }}\left[ \frac{1/a}{(1-S^{2})^{2}}\right] %
\Bigg\{2a^{2}\Gamma ^{2}S^{2}\left[ -1-4S^{2}+F_{0}\left( a^{2}\Gamma
^{2}\right) +4F_{0}\left( \frac{a^{2}\Gamma ^{2}}{4}\right) \right]
\]
\begin{equation}
+2S^2+{{3}\over{2}}S^4-\left(2-{{3}\over{2}}S^2\right)F_0(a^2\Gamma^2)
+2S^{2} \left[ F_{0}\left( \frac{a^{2}\Gamma
^{2}}{4}\right)-\sqrt{S} \right] \Bigg\}, \label{dVda}
\end{equation}


Fig.1 and 2 present the values of the one-body coupling constant for,
 respectively, the non- adiabatic, and the adiabatic limit, evaluated by
 assuming
for the shape-controlling parameter  of the Wannier functions $\Gamma$
the typical\cite{acquarone} value $1.$\AA$^{-1}$.
\begin{figure}
\centerline{\psfig{figure=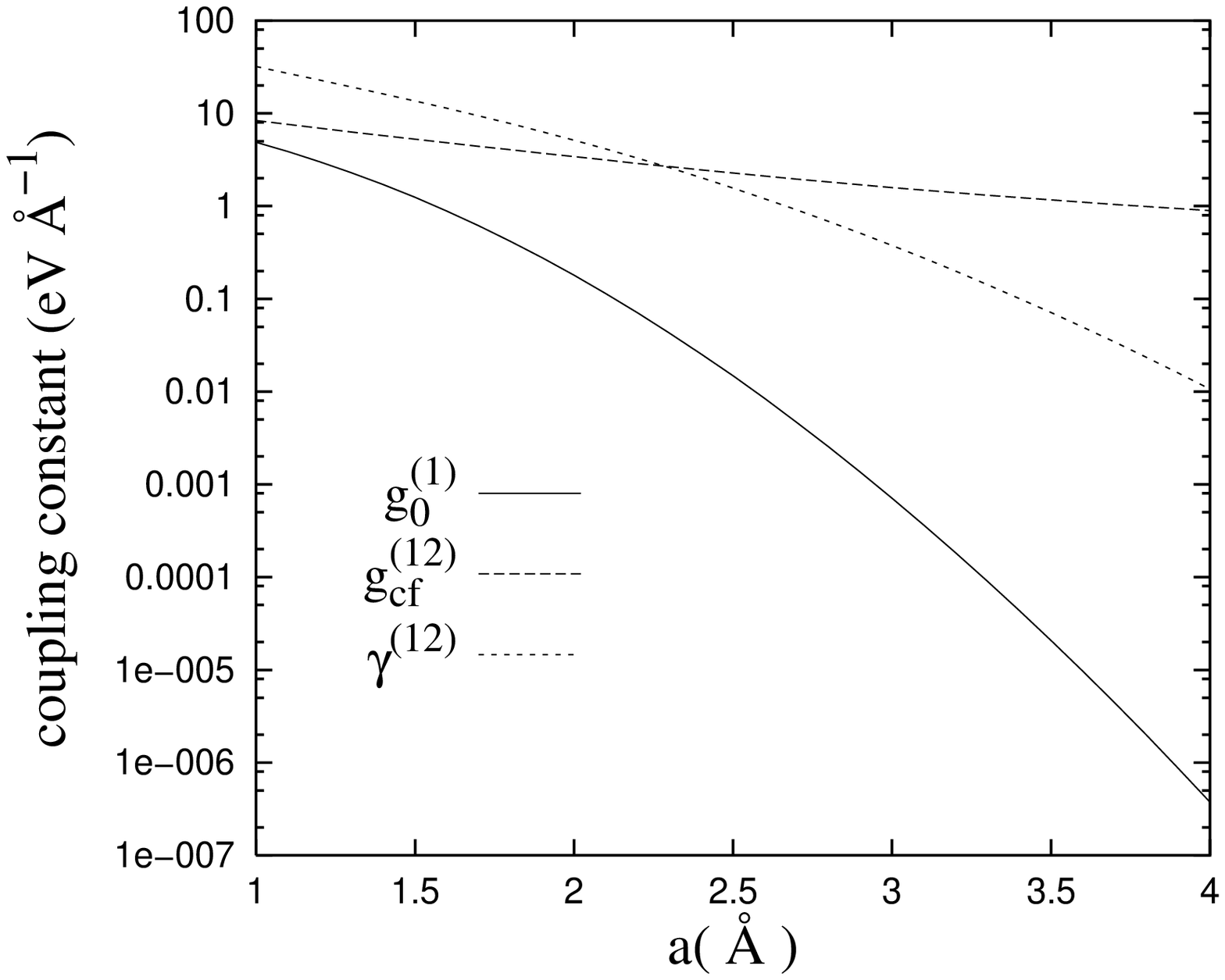,height=6cm,width=8cm}}
\caption {Non-adiabatic coupling constants
$g_0^{(1)},g_{cf}^{(12)},\gamma^{(12)}$ (in eV\AA$^{-1}$) versus the dimer length (in \AA),
evaluated assuming $\Gamma=1.0 \AA^{-1}$.} \label{fig1}
\end{figure}
 The  most
unexpected result concerns $g_{cf}$. While usually neglected in
the literature\cite{barisic} on metallic systems, this coupling
has been recognized as relevant to polar materials\cite{barisic2}.
We find indeed that, when $\Gamma=1.0$\AA$^{-1}$, $g_{cf}$ is
larger than $g_0$ for any $a$,
 and it becomes the largest
parameter for $a>2.2$\AA. For small $a$, the $SSH$ coupling is the
largest.

\begin{figure}
\centerline{\psfig{figure=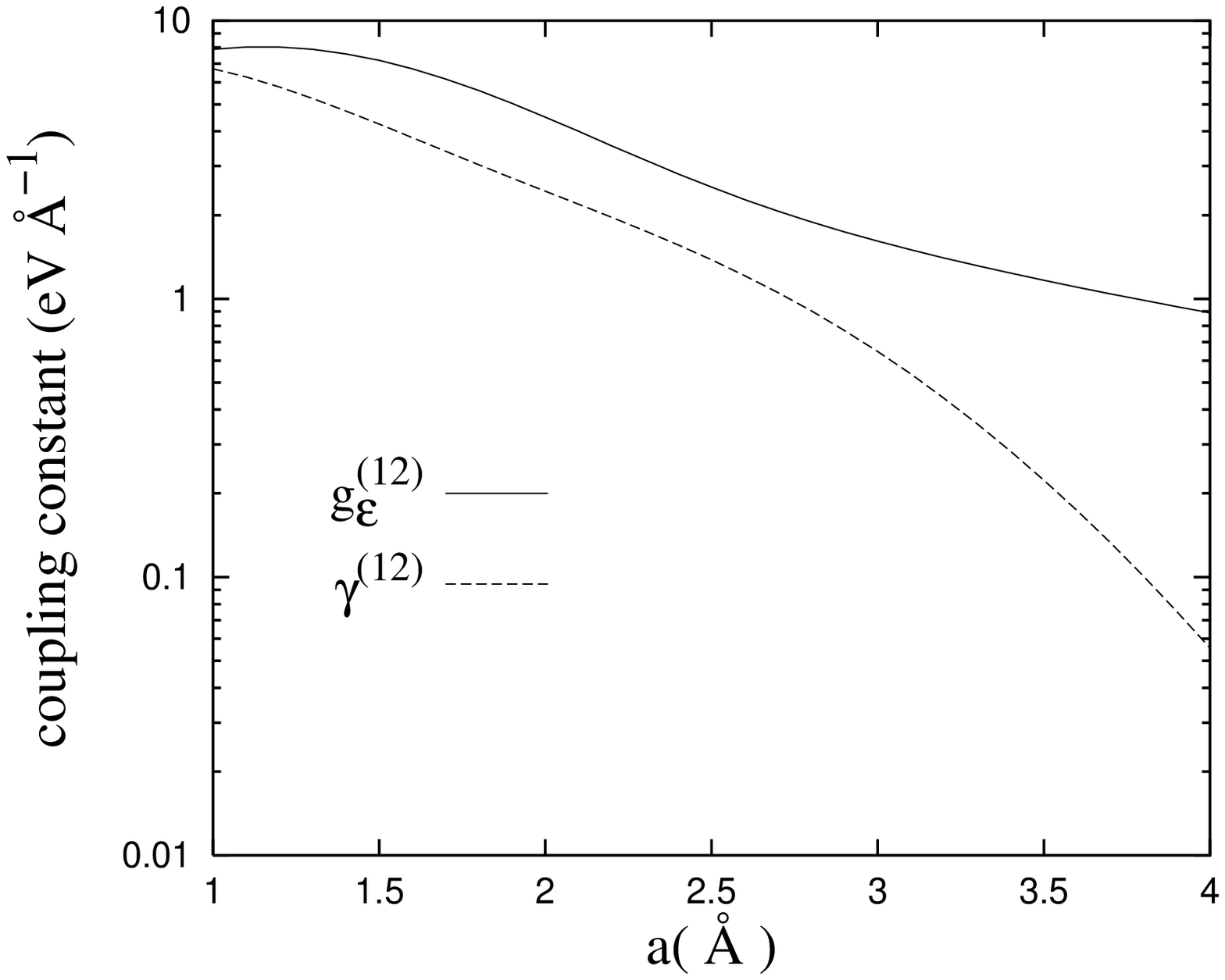,height=6cm,width=8cm}}
\caption {Adiabatic coupling constants
$g_{\epsilon}^{(12)}, \gamma^{(12)}$ (in eV$\AA^{-1}$) versus the
dimer length (in \AA), for $\Gamma=1.0$ \AA$^{-1}$.} \label{fig2}
\end{figure}
Fig.2 for the adiabatic case shows that $g_{\epsilon}^{(12)}$ is always larger
than the $SSH$
interaction $\gamma^{(12)}$, and particularly for large $a$ there is
an order of magnitude difference between them.
\begin{figure}
\centerline{\psfig{figure=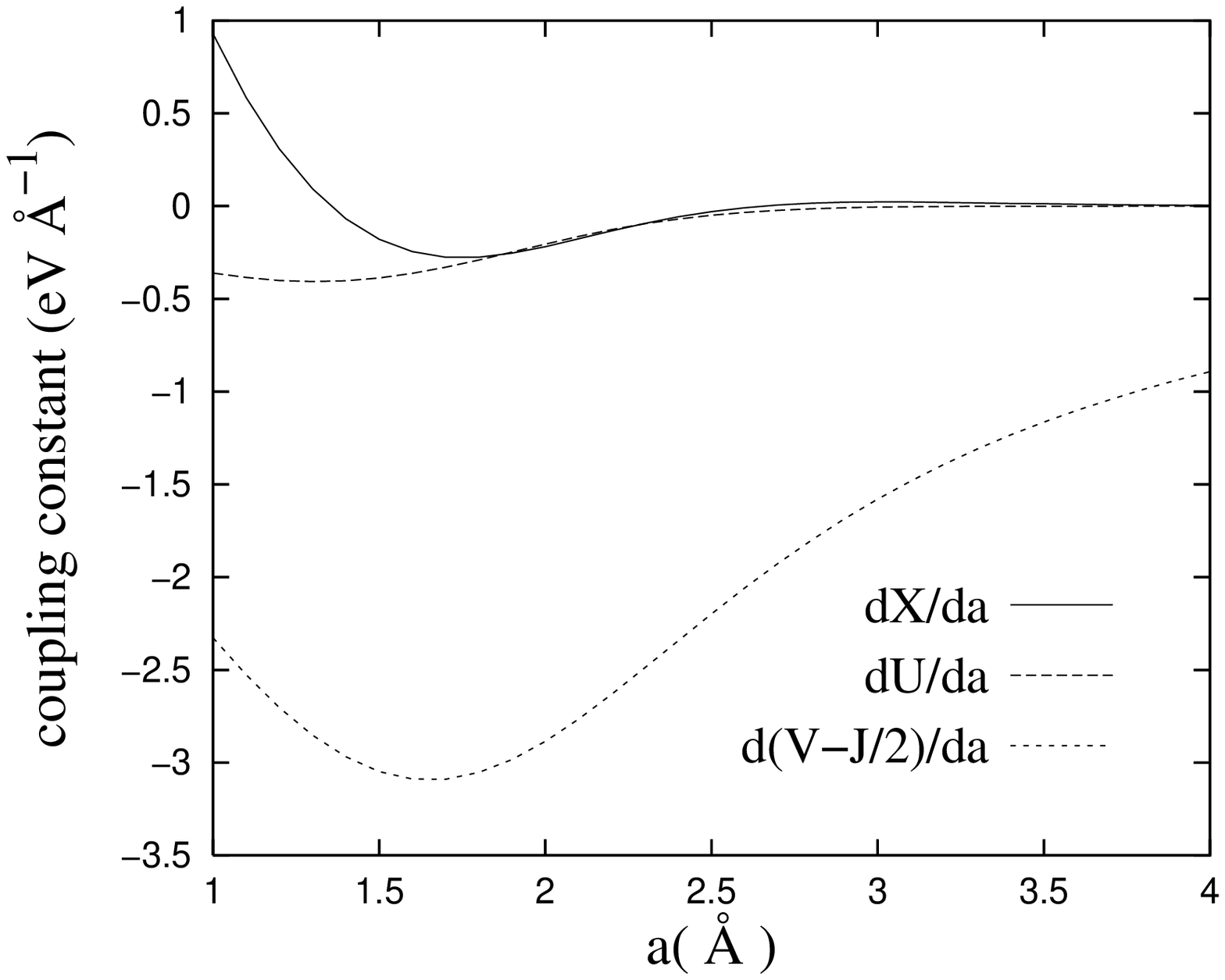,height=6cm,width=8cm}}
\caption {Adiabatic coupling constants from two-body interactions
 $dX/da$, $dU/da$, $d(V-J/2)/da$ (in eV\AA$^{-1}$) versus
the dimer length (in \AA), for $\Gamma=1.0$ \AA$^{-1}$.}
\label{fig3}
\end{figure}
The case of $\Gamma=2.\AA^{-1}$ (discussed in ref.3) shows
that, the more localized
are the orbitals, the more relevant is the role of $g_{cf}$ in
relation to the other admissible couplings.
Fig.3 shows the couplings derived from the two-body electronic
interactions for the same parameters as Fig.2. In general, their
values are much smaller than those of $g_{\epsilon}$ and $\gamma^{(12)}$,
with the possible exception of $d(V-J/2)/da$. Indeed, that coupling
arises from a physical mechanism not very different from the one
originating $g^{(i)}_{V_j}$, i.e. the vibration of the charge on
site $j$ as felt by site $i$. Similarly to $g_{\epsilon}$ also
$d(V-J_z/2)/da$ decreases slowly with $a$, so that for large $a$ those two are
the only relevant couplings. Such interactions in the lattice have been recently discussed in 
ref.12, while their effects in the optical spectra have been treated in ref.13.
\par
\section{Conclusions}

\noindent We have presented the analytical evaluation of the electron-phonon
coupling parameters derived from both one- and two-body electronic interactions
 in a model of a dimer with a non-degenerate orbital
built from atomic orbital of Gaussian shape. We have shown that 
the coupling terms in the adiabatic and the anti-adiabatic limits differ qualitatively.\par 
The evaluation of the coupling terms originating from the two-body
electronic interactions shows that at least the one 
generated by the Coulomb repulsion between the charges on
different sites, is comparable to, or even larger than, the couplings
derived from the one-body interactions. The quantitative
results for  the coupling parameters, even if agreeing in order of
magnitude with some estimates from experimental data\cite{hlubina} 
 are obviously model-depending. However, their ratios
should be more close to the reality. In particular, the obtained
values of the coupling terms, when compared to the values of
the electronic interactions resulting from the same Wannier
functions\cite{acquarone}
 suggests that, for dimer lengths comparable to the lattice parameters
in high temperature superconductors and colossal magnetoresistance materials, 
at most $dU/da$ and $dX/da$ can be safely dropped, 
while neglecting {\it any} of the other
electron-phonon interactions is a questionable approximation. \par

\centerline{Acknowledgements}
\noindent It is a pleasure to
thank J.R. Iglesias, M. A. Gusm\~{a}o, M. Cococcioni,
A. Alexandrov, and particularly A.A. Aligia
and A. Painelli,
for critical discussions and comments. This work was supported by
I.N.F.M. and by MURST 1997 co-funded project "Magnetic Polarons in Manganites".\par


\begin{thebibliography}{99}
\bibitem{noi2} The Hamiltonians of Eqs. \ref{Hnonad} and \ref{Hadiab} differ
 from the
corresponding results in Ref.3, because here we correct some
 mistakes which we found in that paper after it was published.
\end{thebibliography}
\end{document}